\def\sphinxdocclass{article}
\documentclass[a4paper,10pt,english]{sphinxhowto}
\usepackage[utf8]{inputenc}
\DeclareUnicodeCharacter{00A0}{\nobreakspace}
\usepackage{cmap}
\usepackage[T1]{fontenc}
\usepackage{babel}
\usepackage{times}
\usepackage[Bjarne]{fncychap}
\usepackage{longtable}
\usepackage{sphinx}
\usepackage{multirow}

\title{Technical Description of TOBI Interface D (TiD)}
\date{May 01, 2015}
\release{0.3.0.0}
\author{Christian Breitwieser}
\newcommand{\sphinxlogo}{}
\renewcommand{\releasename}{Release}
\makeindex

\makeatletter
\def\PYG@reset{\let\PYG@it=\relax \let\PYG@bf=\relax%
    \let\PYG@ul=\relax \let\PYG@tc=\relax%
    \let\PYG@bc=\relax \let\PYG@ff=\relax}
\def\PYG@tok#1{\csname PYG@tok@#1\endcsname}
\def\PYG@toks#1+{\ifx\relax#1\empty\else%
    \PYG@tok{#1}\expandafter\PYG@toks\fi}
\def\PYG@do#1{\PYG@bc{\PYG@tc{\PYG@ul{%
    \PYG@it{\PYG@bf{\PYG@ff{#1}}}}}}}
\def\PYG#1#2{\PYG@reset\PYG@toks#1+\relax+\PYG@do{#2}}

\expandafter\def\csname PYG@tok@gd\endcsname{\def\PYG@tc##1{\textcolor[rgb]{0.63,0.00,0.00}{##1}}}
\expandafter\def\csname PYG@tok@gu\endcsname{\let\PYG@bf=\textbf\def\PYG@tc##1{\textcolor[rgb]{0.50,0.00,0.50}{##1}}}
\expandafter\def\csname PYG@tok@gt\endcsname{\def\PYG@tc##1{\textcolor[rgb]{0.00,0.27,0.87}{##1}}}
\expandafter\def\csname PYG@tok@gs\endcsname{\let\PYG@bf=\textbf}
\expandafter\def\csname PYG@tok@gr\endcsname{\def\PYG@tc##1{\textcolor[rgb]{1.00,0.00,0.00}{##1}}}
\expandafter\def\csname PYG@tok@cm\endcsname{\let\PYG@it=\textit\def\PYG@tc##1{\textcolor[rgb]{0.25,0.50,0.56}{##1}}}
\expandafter\def\csname PYG@tok@vg\endcsname{\def\PYG@tc##1{\textcolor[rgb]{0.73,0.38,0.84}{##1}}}
\expandafter\def\csname PYG@tok@m\endcsname{\def\PYG@tc##1{\textcolor[rgb]{0.13,0.50,0.31}{##1}}}
\expandafter\def\csname PYG@tok@mh\endcsname{\def\PYG@tc##1{\textcolor[rgb]{0.13,0.50,0.31}{##1}}}
\expandafter\def\csname PYG@tok@cs\endcsname{\def\PYG@tc##1{\textcolor[rgb]{0.25,0.50,0.56}{##1}}\def\PYG@bc##1{\setlength{\fboxsep}{0pt}\colorbox[rgb]{1.00,0.94,0.94}{\strut ##1}}}
\expandafter\def\csname PYG@tok@ge\endcsname{\let\PYG@it=\textit}
\expandafter\def\csname PYG@tok@vc\endcsname{\def\PYG@tc##1{\textcolor[rgb]{0.73,0.38,0.84}{##1}}}
\expandafter\def\csname PYG@tok@il\endcsname{\def\PYG@tc##1{\textcolor[rgb]{0.13,0.50,0.31}{##1}}}
\expandafter\def\csname PYG@tok@go\endcsname{\def\PYG@tc##1{\textcolor[rgb]{0.20,0.20,0.20}{##1}}}
\expandafter\def\csname PYG@tok@cp\endcsname{\def\PYG@tc##1{\textcolor[rgb]{0.00,0.44,0.13}{##1}}}
\expandafter\def\csname PYG@tok@gi\endcsname{\def\PYG@tc##1{\textcolor[rgb]{0.00,0.63,0.00}{##1}}}
\expandafter\def\csname PYG@tok@gh\endcsname{\let\PYG@bf=\textbf\def\PYG@tc##1{\textcolor[rgb]{0.00,0.00,0.50}{##1}}}
\expandafter\def\csname PYG@tok@ni\endcsname{\let\PYG@bf=\textbf\def\PYG@tc##1{\textcolor[rgb]{0.84,0.33,0.22}{##1}}}
\expandafter\def\csname PYG@tok@nl\endcsname{\let\PYG@bf=\textbf\def\PYG@tc##1{\textcolor[rgb]{0.00,0.13,0.44}{##1}}}
\expandafter\def\csname PYG@tok@nn\endcsname{\let\PYG@bf=\textbf\def\PYG@tc##1{\textcolor[rgb]{0.05,0.52,0.71}{##1}}}
\expandafter\def\csname PYG@tok@no\endcsname{\def\PYG@tc##1{\textcolor[rgb]{0.38,0.68,0.84}{##1}}}
\expandafter\def\csname PYG@tok@na\endcsname{\def\PYG@tc##1{\textcolor[rgb]{0.25,0.44,0.63}{##1}}}
\expandafter\def\csname PYG@tok@nb\endcsname{\def\PYG@tc##1{\textcolor[rgb]{0.00,0.44,0.13}{##1}}}
\expandafter\def\csname PYG@tok@nc\endcsname{\let\PYG@bf=\textbf\def\PYG@tc##1{\textcolor[rgb]{0.05,0.52,0.71}{##1}}}
\expandafter\def\csname PYG@tok@nd\endcsname{\let\PYG@bf=\textbf\def\PYG@tc##1{\textcolor[rgb]{0.33,0.33,0.33}{##1}}}
\expandafter\def\csname PYG@tok@ne\endcsname{\def\PYG@tc##1{\textcolor[rgb]{0.00,0.44,0.13}{##1}}}
\expandafter\def\csname PYG@tok@nf\endcsname{\def\PYG@tc##1{\textcolor[rgb]{0.02,0.16,0.49}{##1}}}
\expandafter\def\csname PYG@tok@si\endcsname{\let\PYG@it=\textit\def\PYG@tc##1{\textcolor[rgb]{0.44,0.63,0.82}{##1}}}
\expandafter\def\csname PYG@tok@s2\endcsname{\def\PYG@tc##1{\textcolor[rgb]{0.25,0.44,0.63}{##1}}}
\expandafter\def\csname PYG@tok@vi\endcsname{\def\PYG@tc##1{\textcolor[rgb]{0.73,0.38,0.84}{##1}}}
\expandafter\def\csname PYG@tok@nt\endcsname{\let\PYG@bf=\textbf\def\PYG@tc##1{\textcolor[rgb]{0.02,0.16,0.45}{##1}}}
\expandafter\def\csname PYG@tok@nv\endcsname{\def\PYG@tc##1{\textcolor[rgb]{0.73,0.38,0.84}{##1}}}
\expandafter\def\csname PYG@tok@s1\endcsname{\def\PYG@tc##1{\textcolor[rgb]{0.25,0.44,0.63}{##1}}}
\expandafter\def\csname PYG@tok@gp\endcsname{\let\PYG@bf=\textbf\def\PYG@tc##1{\textcolor[rgb]{0.78,0.36,0.04}{##1}}}
\expandafter\def\csname PYG@tok@sh\endcsname{\def\PYG@tc##1{\textcolor[rgb]{0.25,0.44,0.63}{##1}}}
\expandafter\def\csname PYG@tok@ow\endcsname{\let\PYG@bf=\textbf\def\PYG@tc##1{\textcolor[rgb]{0.00,0.44,0.13}{##1}}}
\expandafter\def\csname PYG@tok@sx\endcsname{\def\PYG@tc##1{\textcolor[rgb]{0.78,0.36,0.04}{##1}}}
\expandafter\def\csname PYG@tok@bp\endcsname{\def\PYG@tc##1{\textcolor[rgb]{0.00,0.44,0.13}{##1}}}
\expandafter\def\csname PYG@tok@c1\endcsname{\let\PYG@it=\textit\def\PYG@tc##1{\textcolor[rgb]{0.25,0.50,0.56}{##1}}}
\expandafter\def\csname PYG@tok@kc\endcsname{\let\PYG@bf=\textbf\def\PYG@tc##1{\textcolor[rgb]{0.00,0.44,0.13}{##1}}}
\expandafter\def\csname PYG@tok@c\endcsname{\let\PYG@it=\textit\def\PYG@tc##1{\textcolor[rgb]{0.25,0.50,0.56}{##1}}}
\expandafter\def\csname PYG@tok@mf\endcsname{\def\PYG@tc##1{\textcolor[rgb]{0.13,0.50,0.31}{##1}}}
\expandafter\def\csname PYG@tok@err\endcsname{\def\PYG@bc##1{\setlength{\fboxsep}{0pt}\fcolorbox[rgb]{1.00,0.00,0.00}{1,1,1}{\strut ##1}}}
\expandafter\def\csname PYG@tok@mb\endcsname{\def\PYG@tc##1{\textcolor[rgb]{0.13,0.50,0.31}{##1}}}
\expandafter\def\csname PYG@tok@ss\endcsname{\def\PYG@tc##1{\textcolor[rgb]{0.32,0.47,0.09}{##1}}}
\expandafter\def\csname PYG@tok@sr\endcsname{\def\PYG@tc##1{\textcolor[rgb]{0.14,0.33,0.53}{##1}}}
\expandafter\def\csname PYG@tok@mo\endcsname{\def\PYG@tc##1{\textcolor[rgb]{0.13,0.50,0.31}{##1}}}
\expandafter\def\csname PYG@tok@kd\endcsname{\let\PYG@bf=\textbf\def\PYG@tc##1{\textcolor[rgb]{0.00,0.44,0.13}{##1}}}
\expandafter\def\csname PYG@tok@mi\endcsname{\def\PYG@tc##1{\textcolor[rgb]{0.13,0.50,0.31}{##1}}}
\expandafter\def\csname PYG@tok@kn\endcsname{\let\PYG@bf=\textbf\def\PYG@tc##1{\textcolor[rgb]{0.00,0.44,0.13}{##1}}}
\expandafter\def\csname PYG@tok@o\endcsname{\def\PYG@tc##1{\textcolor[rgb]{0.40,0.40,0.40}{##1}}}
\expandafter\def\csname PYG@tok@kr\endcsname{\let\PYG@bf=\textbf\def\PYG@tc##1{\textcolor[rgb]{0.00,0.44,0.13}{##1}}}
\expandafter\def\csname PYG@tok@s\endcsname{\def\PYG@tc##1{\textcolor[rgb]{0.25,0.44,0.63}{##1}}}
\expandafter\def\csname PYG@tok@kp\endcsname{\def\PYG@tc##1{\textcolor[rgb]{0.00,0.44,0.13}{##1}}}
\expandafter\def\csname PYG@tok@w\endcsname{\def\PYG@tc##1{\textcolor[rgb]{0.73,0.73,0.73}{##1}}}
\expandafter\def\csname PYG@tok@kt\endcsname{\def\PYG@tc##1{\textcolor[rgb]{0.56,0.13,0.00}{##1}}}
\expandafter\def\csname PYG@tok@sc\endcsname{\def\PYG@tc##1{\textcolor[rgb]{0.25,0.44,0.63}{##1}}}
\expandafter\def\csname PYG@tok@sb\endcsname{\def\PYG@tc##1{\textcolor[rgb]{0.25,0.44,0.63}{##1}}}
\expandafter\def\csname PYG@tok@k\endcsname{\let\PYG@bf=\textbf\def\PYG@tc##1{\textcolor[rgb]{0.00,0.44,0.13}{##1}}}
\expandafter\def\csname PYG@tok@se\endcsname{\let\PYG@bf=\textbf\def\PYG@tc##1{\textcolor[rgb]{0.25,0.44,0.63}{##1}}}
\expandafter\def\csname PYG@tok@sd\endcsname{\let\PYG@it=\textit\def\PYG@tc##1{\textcolor[rgb]{0.25,0.44,0.63}{##1}}}

\def\PYGZus{\char`\_}

\def\PYGZlt{\char`\<}
\def\PYGZgt{\char`\>}

\def\PYGZhy{\char`\-}

\def\PYGZdq{\char`\"}

\makeatother

\begin{document}

\maketitle
\tableofcontents
\phantomsection\label{index::doc}

\section{Introduction}
\label{introduction:introduction}\label{introduction::doc}\label{introduction:tia-documentation-of-tobi-interface-a-version}
TiD (TOBI Interface D) describes a protocol to distribute events and markers used for brain-computer interface (BCI)
purposes. It is based on a client--server principle, whereby the server acts as a distributor,
dispatching incomming messages to every other connected client.
The principle is somehow similar to a bus.

TiD is mainly intended to facilita event distribution in a network envoronment.

It is not intended to replace any direct function calls wihin an established systems.
A direct function call will always be the fastest way to exchange information within on
and the same process, so TiD will never be a replacement for that.

\subsection{License}
\label{introduction:license}
The TiD library is licenced under the \href{http://www.gnu.org/licenses/lgpl.html}{LGPLv3}.

\subsection{Contact}
\label{introduction:contact}
For further information please contact \href{mailto:c.breitwieser@tugraz.at}{c.breitwieser@tugraz.at}.

\section{Design principle}
\label{tcp_connection::doc}\label{tcp_connection:lgplv3}\label{tcp_connection:design-principle}
TiD is designed to distribute BCI events to multiple clients, based on a client--server system.
If a client creates an event, this event is sent to the TiD server, which dispatches it to all
connected clients. To ensure proper timing every TiD message is equipped with a block number
corresponding to the respective block the client was processing when the event occurred. Additional
relative and absolute timestamps (in microseconds) are included into a TiD message, to provide inter-frame
accuracy (assign an event to samples inside a frame).
If a client is not involved in data processing (and is therefore not aware of the actual block number),
the actual frame number is inserted by the TiD server before distributing the event.
Therefore a TiD server needs some communication with the data acquisition system (e.g. TOBI Signal Server
using TiA {\hyperref[tcp_connection:tia-doc]{{[}TiA-Doc{]}}} {\hyperref[tcp_connection:tia-ieee]{{[}TiA-IEEE{]}}}).

To ensure events being always synchronous with the data, every processing module/step should forward
the frame number of the actual data.

\section{Connection principle}
\label{tcp_connection:connection-principle}
A TiD server has to provide a TCP port on which clients can establish a connection. Using an
TCP acceptor, this connection is then bound to a dedicated port on the TiD Server.
Each client gets its own connection to the server.

Via this connection TiD messages are sent to the server and distributed to all other clients.

Some important remarks:
1. The messages are encoded in UTF-8
2. All characters are case sensitive!

\section{TiD Message}
\label{message_format:tid-message}\label{message_format::doc}

\subsection{Structure}
\label{message_format:structure}
Each message, which is send from the client to the server or vice versa, simply contains an XML
TiD message string as follows:

\subsubsection{Example}
\label{message_format:example}
\begin{Verbatim}[commandchars=\\\{\},numbers=left,firstnumber=1,stepnumber=1]
\PYG{n+nt}{\PYGZlt{}tid} \PYG{n+na}{version=}\PYG{l+s}{\PYGZdq{}0.3.0.0\PYGZdq{}}
\PYG{n+na}{description=}\PYG{l+s}{\PYGZdq{}beep\PYGZdq{}}
\PYG{n+na}{block=}\PYG{l+s}{\PYGZdq{}1732\PYGZdq{}}
\PYG{n+na}{family=}\PYG{l+s}{\PYGZdq{}biosig\PYGZdq{}}
\PYG{n+na}{event=}\PYG{l+s}{\PYGZdq{}785\PYGZdq{}}
\PYG{n+na}{absolute=}\PYG{l+s}{\PYGZdq{}1330691458,821096\PYGZdq{}}
\PYG{n+na}{relative=}\PYG{l+s}{\PYGZdq{}34687,761248\PYGZdq{}}
\PYG{n+na}{source=}\PYG{l+s}{\PYGZdq{}P300 detector\PYGZdq{}}
\PYG{n+na}{value=}\PYG{l+s}{\PYGZdq{}3,14159\PYGZdq{}}\PYG{n+nt}{/\PYGZgt{}}
\end{Verbatim}

The TiD message has some mandatory and some optional attributes. These attributes
and their intention are be described in succession.

\subsection{Version}
\label{message_format:version}
\textbf{Mandatory}

A version attribute to avoid version incompatibilities during message processing.

The versions follow the \$CURRENT.\$REVISION.\$MINOR.\$BUGFIX schema and the following rules:
\begin{itemize}
\item {} 
If any ``big'' new features have get added, resulting in heavy interface changes,
increment \$CURRENT, and set all lower fields to ``0''.

\item {} 
If any interfaces have been added, removed, or changed since the last update,
increment \$REVISION, and set all lower fields to ``0''.

\item {} 
If the library source code has changed at all since the last update then
increment \$MINOR and set \$BUGFIX to ``0''

\item {} 
If any bugs have got fixed since the last public release, then increment \$BUGFIX.

\end{itemize}

\subsection{Description}
\label{message_format:description}
\textbf{Mandatory}

A short human readable description of the event.
Mainly intended to distinguish events more easily during manual event inspection
(e.g., ``beep'', ``cue left'', ``flash row 5'', ...)

\subsection{Block}
\label{message_format:block}
The data sample (or potentially block), the event belongs to

Dependent on the data acquisition hardware, a group (or block) of samples might get acquired
together (e.g., done by the g.USBAmp -- \href{http://www.gtect.at}{http://www.gtect.at}).

In such a case, the data acquisition source is acquiring data with a defined sampling rate, but delivers
the data over an API in data blocks. For example, the sampling rare is 500 Hz and the acquisition system
provides data in blocks of 10 samples. Thus, every block contains 10 samples and the blocks are delivered
with a rate of 50 blocks per second.

\subsection{Family}
\label{message_format:family}
\textbf{Mandatory}

The family can be seen as a parent group, the event belongs to.
For example, the biosg project (\href{http://biosig.sf.net}{http://biosig.sf.net}) already defines a big amount of unique events
for different areas (\href{http://sourceforge.net/p/biosig/code/ci/master/tree/biosig4matlab/doc/eventcodes.txt}{http://sourceforge.net/p/biosig/code/ci/master/tree/biosig4matlab/doc/eventcodes.txt}).

In a common environment, all events might usually be from the same family, so the occurrence of an event
clash or event misinterpretation is unlikely. However, to avoid such issues, an event family can get defined in TiD.

That way, a client can choose to react only on events from a known family. So even heterogeneous systems
with different event sources become possible.

Current event family definitions available in TiD:
\begin{itemize}
\item {} 
biosig \emph{(to support events defined by the biosig project)}

\item {} 
custom \emph{(to support custom events, not being defined anywhere)}

\end{itemize}

The number of event families could also get extended in future to further support events fro other
prominent BCI systems, as mentioned below.

Potential other families for the future:
\begin{itemize}
\item {} 
BCI2000

\item {} 
OpenViBE

\end{itemize}

\subsection{Event}
\label{message_format:event}
\textbf{Mandatory}

In TiD an event is treated as a an occurence of a unique happening. Out of this reasons, events
are currently assigned to an event code, inspired by the biosig project, mentioned above.

The event is just an integer value.
This brings the advantage of fast event processing without the need to parse event strings.

In reverse, every event type (for example a ``beep'') needs to get an event code assigned.
It is a valid point of discussion to provide simple strings as a potential event as well.
The current implementation only provides the processing of integer values.

\subsection{Absolute}
\label{message_format:absolute}
\textbf{Mandatory}

An absolute timestamp in microseconds since 1970-01-01 00:00:00.
Internally, the ``gettimeofday()'' method is utilized to obtain the time.

With this value, a proper synchronization of events becomes possible.
Please note: TiD does not offer a clock synchronization mechanism itself. The system
clocks of different computers need to get synchronized by other systems like
NTP (network time protocol) or PTP (precision time protocol).

\subsection{Relative}
\label{message_format:relative}
\textbf{Mandatory}

This timestamp provides a timing value in microseconds, relative to a definable time point.
Currently, the start of the TiD server is used as a reference.

That way, a customizable point of time can get used as ``0'' and a dedicated timeserver as stratum.

Furthermore, the relative timestamp is also allowed to get reset at any point of time, for
example to trace the processing pipeline. By resetting the relative timestamp when the event is created
and reading out its value when the event is finally processed, tracing becomes easily possible, especially
on local systems.

\subsection{Source}
\label{message_format:source}
\textbf{Optional}

The source provides an optional field to specify the event origin like a P300 speller, etc.
It is a simple descriptive string, which could be used to process events only from a defined source.

\subsection{Value}
\label{message_format:value}
\textbf{Optional}

The value is an optional floating point number, defining a potential value related to the event.
For example, ``BeepWithFrequency'' as event and ``1500'' as value.

A more generic value typing system (e.g., to also allow sting) would be a meaningful extension.
However, it was just not needed up to now.

\section{TiD Server}
\label{tid_client_server::doc}\label{tid_client_server:tid-server}
The server is responsible for receiving and dispatching of incoming events.

Events without a frame number and/or timestamp get the actual values from the data acquisition.
An unknown block number is marked with ``-1''.
Therefore it needs to have access to the data acquisition system. It is suggested to include a
TiD server into the data acquisition like the TOBI SignalServer.

It provides the possibility to gather events received from connected clients for saving purposes.
These events are stored an can get saved elsewhere (e.g. in a simple text file or a .gdf file).

All TiD connections are running in separated threads to facilitate fast event processing, especially
on nowadays multi-core computer systems. In case of an error, the connection and the receiver thread
get closed.

The TiD Server is listening at port 9001 by default.
To decrease delivery latency, Nagle's algorithm is deactivated (TCP\_NODELAY flag is set).

\section{TiD client}
\label{tid_client_server:tid-client}
A TiD client can send and receive TiD messages using the connection to the TiD server. All
communication is handled via this connection.

The client is responsible by itself for processing the TiD messages. It can simply ignore messages
it does not care for.

Every client, also processing incoming data (e.g. biosignals, classification results) has to include
the according frame number into an outgoing TiD message.
If a client is not in touch with the BCI processing chain (pre-processing, feature extraction,
classification, fusion, shared control), e.g. a P300 speller just sending events, has to leave the
frame number blank, the actual value will be inserted by the server before dispatching.

\section{TiD Architecture}
\label{tid_client_server:tid-architecture}
As mentioned at the beginning, TiD is acting in a bus oriented manner.
The following figure illustrates the principle:
\begin{figure}[htbp]
\centering

\scalebox{0.900000}{\includegraphics{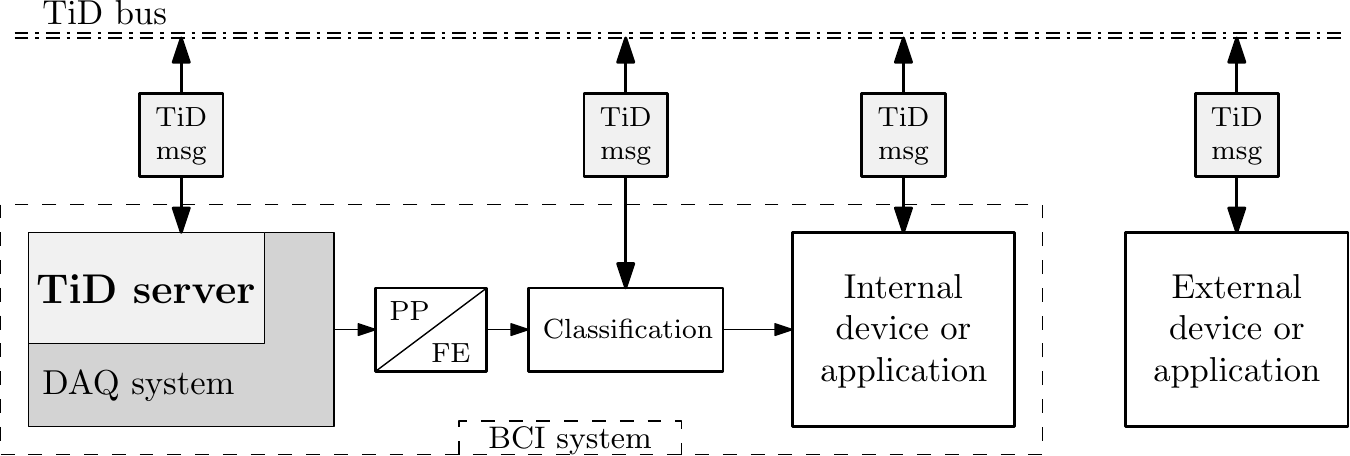}}
\end{figure}

\section{TiD Performance Tuning}
\label{tid_client_server:tid-performance-tuning}

\subsection{Unix/Linux:}
\label{tid_client_server:unix-linux}
In a Unix/Linux environment it is suggested to use a custom kernel.

Potential tuning options would be the modification of the timer frequency:
frequency to quickly react on events

\begin{Verbatim}[commandchars=\\\{\}]
Symbol: HZ\PYGZus{}1000 [=y]
Location:
\PYGZhy{}\PYGZgt{} Processor type and features
\PYGZhy{}\PYGZgt{} Timer frequency (\PYGZlt{}choice\PYGZgt{} [=y])
\end{Verbatim}

And the modification of the preemption model:

\begin{Verbatim}[commandchars=\\\{\}]
Symbol: PREEMPT\PYGZus{}VOLUNTARY [=n]
Location:
\PYGZhy{}\PYGZgt{} Processor type and features
\PYGZhy{}\PYGZgt{} Preemption Model (\PYGZlt{}choice\PYGZgt{} [=y])
\end{Verbatim}

As suggested by RedHat (\href{http://developerblog.redhat.com/2015/02/11/low-latency-performance-tuning-rhel-7/}{http://developerblog.redhat.com/2015/02/11/low-latency-performance-tuning-rhel-7/}, \href{https://access.redhat.com/videos/898583}{https://access.redhat.com/videos/898583}),
adjusting system settings can also affect network latency.

According to the RedHat sources mentioned above following additional settings for sysctl.conf should be meaningful

\begin{Verbatim}[commandchars=\\\{\}]
net.core.wmem\PYGZus{}max=12582912
net.core.rmem\PYGZus{}max=12582912
net.ipv4.tcp\PYGZus{}rmem= 10240 87380 12582912
net.ipv4.tcp\PYGZus{}wmem= 10240 87380 12582912


net.core.busy\PYGZus{}read=50
net.core.busy\PYGZus{}pull=50
net.ipv4.tcp\PYGZus{}fastopen=1
kernel.numa\PYGZus{}balancing=0

kernel.sched\PYGZus{}min\PYGZus{}granularity\PYGZus{}ns = 10000000
kernel.sched\PYGZus{}wakeup\PYGZus{}granularity\PYGZus{}ns = 10000000
vm.dirty\PYGZus{}ratio = 10
vm.dirty\PYGZus{}background\PYGZus{}ratio = 3
vm.swappiness=10
transparent\PYGZus{}hugepage=never
kernel.sched\PYGZus{}migration\PYGZus{}cost\PYGZus{}ns = 5000000
\end{Verbatim}

\subsection{Windows:}
\label{tid_client_server:windows}
The windows scheduler acts, depending on operating system and host related configuration, in defined
time slots, also often called ``quantum''.
Such a quantum can get interpreted as a guaranteed amount of time for a certain thread/process.

Such quantums can get interrupted by operating system interrupts. As discussed in ``The Windows Timestamp Project''
(\href{http://www.windowstimestamp.com/description}{http://www.windowstimestamp.com/description}), Windows offers some undocumented functions to reconfigure the
interrupt time resolution (like ``\emph{NtSetTimerResolution}'').

The code within the performance tests of the TiD library already contains some demo code, setting
the scheduler to the maximum timer granularity, resulting in potential smaller delays, even in higher
workload situations.

Unfortunately Microsoft does not offer as many latency tuning options as Linux. Some additional
options, which are unfortunately (partly) only available on Windows Server OS are described on the following iste:
\href{https://technet.microsoft.com/en-us/library/jj574151.aspx}{https://technet.microsoft.com/en-us/library/jj574151.aspx}

Thus, it is suggested to disable power saving options of the network adapter and further also disable
power saving options from the CPU (like C-states).

\section{TiD Performance Tests}
\label{tid_client_server:tid-performance-tests}
The TiD library offers automated performance testing methods.
These methods are available in the ``tid\_tests'' sub-project.

Within these tests, the individual latency then sending/receiving or dispatching a defined amount
of messages for variable TiD message lengths can get obtained.
Furthermore, the testing of the localhost as well as the network latency is possible.

This sub-project also offers Matlab script to load and analyze the test results. Histograms show the overall network latency. The individual transfer function for the individual
plots were calculated as well, presenting the influence of the jitter when averaging data series, aligned
on the basis of TiD events.

That way, every TiD user can determine the network environment latency conditions on his/her own to get a
feeling for the potential limitations.

\renewcommand{\indexname}{Index}
\printindex

\begin{thebibliography}{TiA-IEEE}
\bibitem[TiA-Doc]{TiA-Doc}{\phantomsection\label{tcp_connection:tia-doc} 
C. Breitwieser and C. Eibel, ``TiA -- Documentation of TOBI Interface A'',
ArXiv e-prints, Mar. 2011. (\href{http://arxiv.org/abs/1103.4717}{http://arxiv.org/abs/1103.4717})
}
\bibitem[TiA-IEEE]{TiA-IEEE}{\phantomsection\label{tcp_connection:tia-ieee} 
C. Breitwieser, I. Daly, C. Neuper, and G. R. M?ller-Putz, ``Proposing a standardized protocol for raw biosignal transmission'', IEEE. Trans.
Biomed. Eng., vol. 59, no. 3, pp. 852-859, 2012.
}
\end{thebibliography}
\end{document}